\documentstyle[12pt,epsf]{article}

\newcommand{\beq}{\begin{equation}}
\newcommand{\eeq}{\end{equation}}
\newcommand{\beqa}{\begin{eqnarray}}
\newcommand{\eeqa}{\end{eqnarray}}
\newcommand{\dfrac}{\displaystyle \frac}

\begin{document}

\begin{center}
{\large\bf ON THE INTERPRETATION OF $\pi N \to
 \pi\pi N$ DATA \\  NEAR THRESHOLD}
\end{center}

\vspace{.5in}

\begin{center}
M.G. Olsson$^a$\footnote{electronic
address: olsson@phenxg.physics.wisc.edu},
Ulf-G. Mei{\ss}ner$^b$\footnote{electronic address:
meissner@pythia.itkp.uni-bonn.de},
N. Kaiser$^c$\footnote{electronic address: nkaiser@physik.tu-muenchen.de},
V. Bernard$^d$\footnote{electronic address: bernard@crnhp4.in2p3.fr}\\

\bigskip

\bigskip

%\bigskip

\it
$^a$Department of Physics, University of Wisconsin, Madison, WI 53706, USA\\
$^b$Universit\"at Bonn, Inst. f{\"u}r Theoretische Kernphysik,
D-53115 Bonn, FRG\\
$^c$TU M\"unchen, Physik Department T39, D-85747 Garching, FRG\\
$^d$CRN Strasbourg, Physique Th\'eorique, F-67037 Strasbourg Cedex 2, France\\
\end{center}

\vspace{1in}

\thispagestyle{empty}

\begin{abstract}
Near threshold pion production experiments have been recently carried out
and used to extract S--wave $\pi\pi$ scattering lengths.
We emphasize here that at present these processes are related only at
the tree level (and its first correction) in chiral
perturbation theory. Higher order corrections (including loops)
must be evaluated before
rigorous claims concerning S--wave $\pi\pi$ scattering lengths can be made.
\end{abstract}

\vfill

\noindent CRN 95-13

\noindent MADPH-95-866

\noindent TK 95 07 \hfill March 1995

\newpage

\section{Introduction}
\label{sec:intro}
%%%
In the last few years an impressive series of experiments have measured
the total cross section for the processes  $\pi N\to\pi\pi N$ quite close to
threshold~\cite{kernel,pocanic,kernel2,kernel3,lowe}.  The modulus of the
threshold amplitude is then found by the extrapolation
\begin{equation}
| {\cal A}(\pi\pi N) |^2 = \lim_{T_\pi\to T_\pi^{th}}
{ \sigma(\pi N\to\pi\pi N) \over C \, S \,(T_\pi-T_\pi^{th})^2  }
                                                  \end{equation}
where $T_\pi$ is the incident laboratory pion kinetic energy, $S$ is a Bose
symmetry factor ($S=1/2$ if the final two pions are identical, otherwise
it is unity), and
\begin{equation}
      C = M_\pi^2 \, \left(1\over128\pi^2\right) \sqrt3\, (2+\mu)^{1/2}\,
      (2+3\mu)^{1/2} \, (1+2\mu)^{-11/2}
%      C = \left(g_{\pi N}^2\over4\pi\right) 2\sqrt3 (1+2\mu)^{-5/2}
%\left(1+{1\over2}\mu\right)^{-5/2} \left(1+{3\over2}\mu\right)^{-1/2}
\end{equation}
where  $\mu=M_\pi/m$ the ratio of the pion to nucleon mass.
The threshold modulus has been obtained in this way for the five charge states
initiated by $\pi^\pm p$~\cite{burkhard}.  Explicit isospin violation due to
the electromagnetic mass differences has been removed through the kinematics of
the threshold $T_\pi^{th}$ value and the threshold amplitude modulus is assumed
to be isospin invariant.

By Watson's theorem~\cite{watson} the threshold amplitude has the phase of the
initial elastic $J^P={1\over2}^+$ amplitude (up to an overall sign).  The
threshold production amplitude complex phase is then
$\delta_{31}\simeq-4^\circ$ for initial isospin 3/2 and
$\delta_{11}\simeq2^\circ$
for total isospin 1/2.  The threshold production amplitude is thus nearly
real.  At threshold the final $\pi\pi$ state must have isospin 0 or 2 by
extended Bose symmetry and hence there are only two independent threshold
amplitudes; called ${\cal A}_{2I , I_{\pi\pi}}$ (with $I$ the total
isospin of the incident $\pi N$ system and $I_{\pi \pi}$ the isospin of
the two--pion system in the final state).

\bigskip
%\begin{table}
%\caption{}
\def\arraystretch{1.5}
\arraycolsep=1em
\[\begin{array}{r@{\to}lcc}
\hline\hline
\multicolumn{2}{c}{\rm process}\\ \noalign{\vskip-8pt}
\multicolumn{2}{c}{\rm amplitude}
& {\cal A}_{32}(\pi\pi N)&{\cal A}_{10}(\pi\pi N)\\ \hline
\pi^+p& \pi^+\pi^+n& {2\over\sqrt5}& 0\\
&\pi^+\pi^0p& -{1\over\sqrt{10}} &0\\
\pi^-p& \pi^+\pi^-n& {1\over 3\sqrt5}& -{\sqrt2\over3}\\
&\pi^0\pi^0n& {2\over 3\sqrt5}& {\sqrt2\over3}\\
&\pi^-\pi^0p& -{1\over\sqrt{10}} & 0\\
\hline\hline
\end{array}\]
%\end{table}
\smallskip

\centerline{Table 1: Clebsch-Gordan coefficients}

\bigskip

In Table~1
we list the five measured process amplitudes in terms of the two independent
isospin amplitudes ${\cal A}_{32}(\pi\pi N)$ and  ${\cal
  A}_{10}(\pi\pi N)$. From the
measured process amplitude moduli a unique value of ${\cal A}_{10}$
 and ${\cal A}_{32}$ can be found up to an overall sign.

\section{Relation to $\pi \pi$ scattering}
\label{sec:pipi}
%%%
The purest process to test the chiral dynamics of QCD is the reaction
$\pi \pi \to \pi \pi$ in the threshold region. The pertinent partial
waves admit an energy expansion of the type
\beq
t_l^I (s) = q^{2l} \, \lbrace a_l^I + q^2 b_l^I + \ldots \rbrace
\label{pw}
\eeq
with $q$ the modulus of the
 pion three--momentum, $s=4(M_\pi^2 + q^2)$ the cms energy
squared and $l \, (I)$ denote the angular momentum (isospin) of the
$\pi \pi$ system. As first pointed out by Weinberg \cite{weinberg},
the $\pi \pi$ scattering amplitude can be written in terms of one
invariant function $A(s,t,u)$ which takes the form
\beq
A(s,t,u) = \frac{s - M_\pi^2}{F_\pi^2} + {\cal O}(E^4)
\label{a2}
\eeq
where $F_\pi \simeq 93$~MeV is the pion decay constant and ${\cal
  O}(E^4) = {\cal O}(s^2, sM_\pi^2,M_\pi^4,\ldots)$ are corrections which can
not be calculated from current algebra. Consequently,
the S--wave scattering lengths
\beq
a_0^0 (\pi \pi)\, = \frac{7 M_\pi^2}{32 \pi F_\pi^2} \, ,
\quad a_0^2 (\pi \pi)\,  = -\frac{2 M_\pi^2}{32 \pi F_\pi^2} \, ,
\label{as}
\eeq
vanish in the chiral limit, $M_\pi \to 0$, and are
therefore particularly sensitive to the explicit chiral symmetry
breaking in QCD. Furthermore, the one--loop corrections to the Weinberg
result have been worked out \cite{gl83} and rather accurate
predictions could be given, i.e. $a_0^0 (\pi \pi) = 0.20 \pm 0.01$
 \cite{gl84}. It is, however, not straightforward to determine
these fundamental quantities experimentally. Therefore, any option to do this
 is highly welcome (for a review, see e.g. \cite{ugm}).
In what follows, we will be
concerned with one possible candidate, namely the reaction $\pi N \to
\pi \pi N$ at threshold.\footnote{Here, we are not considering
$\pi N \to \pi \pi N$ data at
higher  energy which might be analyzed with the help of Chew--Low
type techniques to give the $\pi \pi$ phases.}

\bigskip

{}From the effective Lagrangian formulation~\cite{weinberg} of PCAC and the
algebra of currents Olsson and Turner (OT)~\cite{olleaf} showed in effect that
\begin{equation}
\def\arraystretch{1.5}
\begin{array}{rcl}
   {\cal A}_{32}(\pi\pi N) &=& -2\sqrt{10} \pi \dfrac{g_{\pi N}}{m} \,
   \Bigl[ \dfrac{a^2_0(\pi\pi)}{M_\pi^2} + d_2 \Bigr] \\
   {\cal A}_{10}(\pi\pi N) &=& \phantom+ 4 \pi \dfrac{g_{\pi N}}{m} \,
    \Bigl[\dfrac{a_0^0(\pi\pi)}{M_\pi^2} + d_0\bigr]
\end{array}
\label{a_I}
\end{equation}
with $g_{\pi  N} =13.4$ the strong pion--nucleon coupling constant.
The above
result is a consequence of the dominance of the pion exchange and contact
diagrams  shown in Fig.~1(a),1(b). To lowest order,
the two ``shift" constants $d_I$ arise from the
sub-leading diagrams of Fig.~1(c). The $d_I$ are of order ${\cal O}(M_\pi)$.
A remark on the diagram 1(a) is in
order here.  One frequently finds in the literature the
erroneous statement that the threshold $\pi N \to \pi \pi N$ amplitudes can
{\it not} be directly related to the $\pi \pi$ phase shifts since the
exchanged pion in the pion pole graph is off mass--shell. However, a general
argument invoking only unitarity tells us that the residue of the pion--pole
term must factor into  the product of the {\it on-shell} $\pi \pi$ scattering
amplitude times the pion--nucleon vertex function (in the $t$--channel)
\cite{dawei}.

\bigskip

\bigskip

\hskip 1in
\epsfxsize=4in
\epsfysize=1in
\epsffile{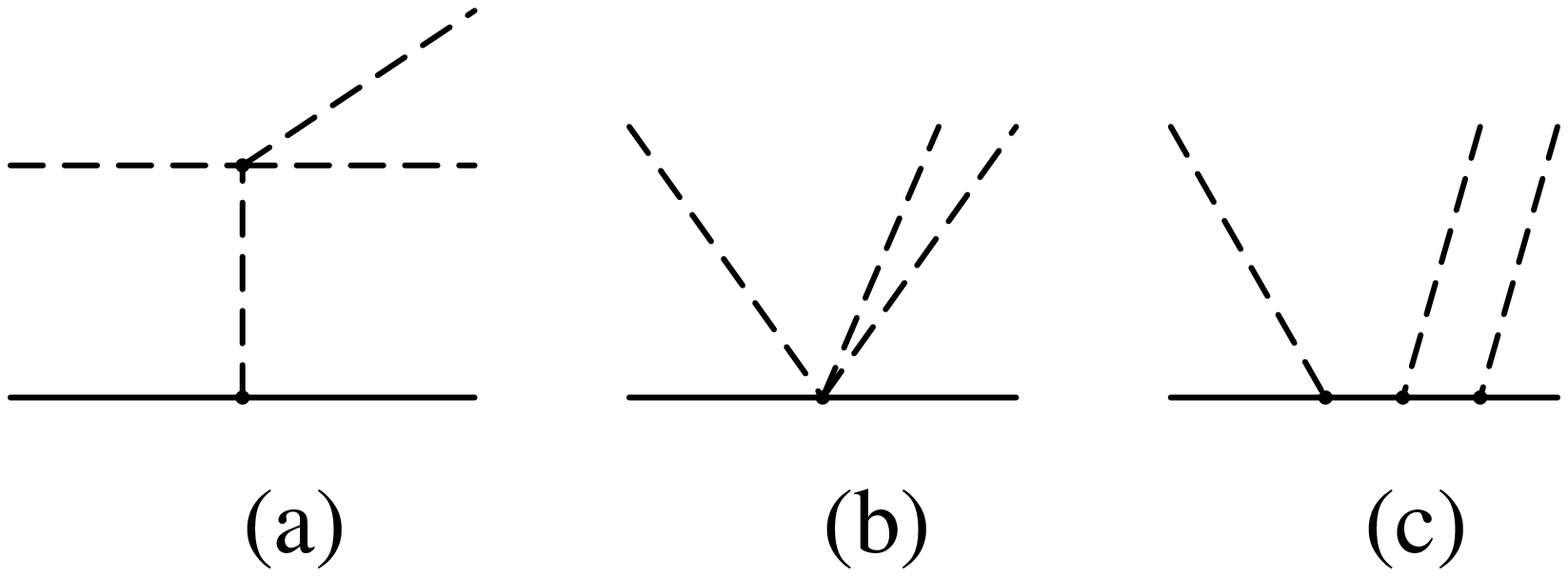}

\bigskip

{\noindent\narrower \it Fig.~1:\quad  Pion pole (a) and contact (b)
  diagrams which lead to the OT relation (\ref{a_I}). The consecutive
pion emission (c) contributes to the shift constants $d_{0,2}$.

\smallskip}
%\vskip -0.5truecm

\bigskip

The OT relation (\ref{a_I}) predates QCD and its
expression through chiral perturbation theory (ChPT). Nevertheless, the OT
production amplitude is equivalent  to the tree level ChPT result
if the $a_0^I(\pi\pi)$  are the tree level $\pi\pi$ S--wave scattering lengths.
In the original formulation, the OT relation contains a parameter
called $\xi$. Its meaning and relevance for present day
data analysis is discussed in the next section.

\section{The $\xi$ parameter}
\label{sec:xi}
%%%

Let us elaborate on the OT $\xi$ parameter \cite{olleaf}. $\xi$
described the pattern of chiral symmetry breaking in the pre QCD era
of the effective Lagrangian.  Only $\xi=0$ is consistent with QCD. To see
this in more detail, consider the so--called $\sigma$--commutator, i.e.
the commutator between an axial charge $Q_5^a$ and
the divergence of the axial current, $D^b = \partial^\mu A^b_\mu$,
\begin{equation}
i \, [Q_5^a \, , \, D^b] = \sigma^{ab} = - F_\pi M_\pi^2 \,
\biggl\lbrace \delta^{ab} \bigl(F_\pi - \frac{\mbox{\boldmath$\pi$}^2}{2F_\pi}
\bigr) +
\frac{\xi}{4F_\pi} \bigl( \delta^{ab} + 2 \pi^a \pi^b \bigr) \biggr\rbrace
\label{sigma}
\end{equation}
where the first term is an isoscalar and the second an isotensor.
The corresponding $\pi \pi$ scattering amplitude to lowest order is
then given by
\beq
A_\xi (s,t,u) = \frac{1}{F_\pi^2} \biggl[ s - M_\pi^2 (1+
\frac{\xi}{2} ) \biggr] \, \, ,
\label{a2xi}
\eeq
and the scattering lengths $a_0^0 (\pi \pi)$ and $a_0^2 (\pi \pi)$ depend
on $\xi$,
\beq
 a_0^0 (\pi \pi)\, = \frac{ M_\pi^2}{32 \pi F_\pi^2} ( 7 -\frac{5}{2}\xi ) \, ,
\quad a_0^2(\pi \pi)\,  = -\frac{M_\pi^2}{32 \pi F_\pi^2} (2 + \xi)\, .
\label{asxi}
\eeq
These forms are frequently used in the literature \cite{pocanic}
 \cite{burkhard} to determine the S--wave $\pi \pi$ scattering
 lengths from the measured and extrapolated $\pi N \to \pi \pi N$ data.
However, in QCD,
the $\sigma$--commutator stems from the explicit chiral symmetry breaking
quark mass term, i.e. (in the isospin limit $m_u = m_d = \hat{m}$)
\begin{equation}
 \sigma^{ab} = \delta^{ab} \, \hat{m} \, ( \bar u u + \bar d d )
\label{sigmaQCD}
\end{equation}
which is purely isoscalar and thus $\xi_{\rm QCD}= 0$.
It is worth to stress that
$\xi =0$ also holds in the so--called `generalized ChPT' \cite{orsay,stern}.
In that scheme, the symmetry--breaking terms are subject to another
counting\footnote{While in standard ChPT, one has $m_s/B \ll 1$, in
  GChPT one assumes $m_s \simeq B$, with $B$  the order parameter
 of the chiral symmetry breaking, $B=-<0|\bar qq|0>/F_\pi^2$.}
which for example modifies even the lowest order (Weinberg) expression for the
elastic $\pi \pi$ scattering amplitude,
\beq
A_{\rm GChPT} (s,t,u) = \frac{1}{F_\pi^2} \biggl[ s - M_\pi^2 (1 -
\frac{\chi}{3} ) \biggr] \, + {\cal O}(E^3) \, ,
\label{a2gchpt}
\eeq
Notice that the corrections start at order $E^3$ in contrast to the
standard scenario, cf. eq.(\ref{a2}). The new parameter $\chi$  measures
the deviation from the conventionally
 adopted (and presumably correct) quark mass
ratio $m_s / \hat{m} = 2M^2_K/M_\pi^2 - 1 \simeq 25$, i.e.
\beq
\chi = 6 \frac{(2M_K^2/M_\pi^2-1) -(m_s/\hat{m})}{ (m_s/\hat{m})^2 -1} \,
\, ,
\eeq
neglecting the small OZI violation in the $0^{++}$ channel. Nevertheless, the
explicit symmetry breaking is still purely isoscalar, i.e. $\xi = 0$.

\section{An improved low--energy representation}
\label{sec:bkm}
%%%

We observe from (\ref{a_I}) that the effect of the constants $d_I$ is to shift
the $\pi\pi$ scattering length values relative to the measured production
amplitudes.
It is crucial therefore to reliably estimate their values.  We point out here
that the loop corrections, counter terms, and other contributions of higher
order ChPT will alter both the scattering lengths and the $\pi \pi N$
threshold amplitudes.

The leading order ChPT corrections~\cite{bernard} are shown in
Fig.~2. Other corrections which must be considered in the shift parameters
involve intermediate $\Delta_{33}$ and possibly other resonances in diagrams
similar to Fig.~1(c). Such tree contributions are implicitly
contained e.g.\ in the model of Oset and Vicente--Vacas \cite{oset}
which is intended to describe the data over a wide range of energies.

In ref.\cite{bernard}, the corrections to the OT relation of order
$M_\pi$ were worked out. This leads to an improved low--energy
representation of the form\footnote{Notice that the overall sign
of the ${\cal A}_{10,32}$ at threshold is fixed by the chiral expansion.}
\begin{equation}
\def\arraystretch{1.5}
\begin{array}{rcl}
   {\cal A}_{32}(\pi\pi N) &=& -2 \sqrt{10} \pi \, \bigl( 1 + \frac{7}{2} \mu
   \bigr) \,   \Bigl[ \dfrac{a_0^2(\pi\pi)}{M_\pi^2} + \tilde{d}_2
                      M_\pi^2  \Bigr] \\
   {\cal A}_{10}(\pi\pi N) &=& \phantom+ 4 \pi \, \bigl( 1 + \frac{37}{14} \mu
   \bigr) \Bigl[\dfrac{a_0^0(\pi\pi)}{M_\pi^2} +
                    \tilde{d}_0 M_\pi^2  \Bigr]
\end{array}
\label{a_Inew}
\end{equation}
where the new shift constants $\tilde{d}_{0,2}$ have the form
\beq
\tilde{d}_I = c_I^0  \, + \, c_I^1 M_\pi  \, + \, c_I^2 M_\pi^2 \, +
\ldots \, , \quad I = 0 , 2
\label{shift}
\eeq
modulo logs. One notices that the correction of order $M_\pi$ is
comparable in size to the leading term (approximately
40$\%$ and 50$\%$ for ${\cal A}_{10}$ and ${\cal A}_{32}$, respectively).
Therefore, it is mandatory to
calculate (at least) the coefficients $c_I^0$. Also, at that order the
one--loop corrections to the S--wave $\pi \pi$ scattering lengths
appear \cite{gl83,gl84}.
One can, however, estimate the ${\cal O}(M_\pi^2)$ corrections
by calculating the unambiguous absorptive parts of the one--loop
diagrams \cite{bernard}.
The corresponding corrections are small for ${\cal A}_{32} (\pi
\pi N)$ and of the order of 30$\%$ for ${\cal A}_{10} (\pi \pi N)$. Such a
pattern is expected, the $\pi \pi$ interactions are small for $I =2$
but sizeable for $I = 0$. Such estimates should only be considered
indicative and can not substitute for the complete calculation of the
shift constants $\tilde{d}_I$.

\bigskip

\bigskip

\hskip 1in
\epsfxsize=4in
\epsfysize=1.1in
\epsffile{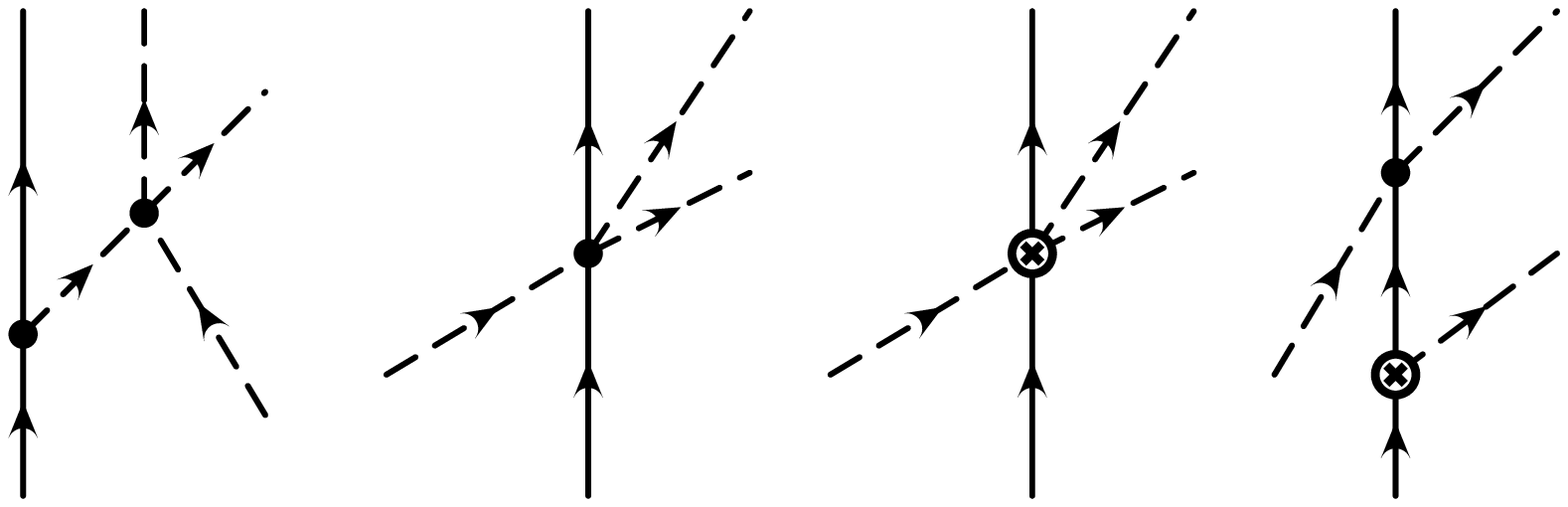}

\bigskip

{\noindent\narrower \it Fig.~2:\quad  Diagrams which give the
  contributions to ${\cal A}_{10} (\pi \pi N)$ and ${\cal A}_{32} (\pi \pi N)$
up-to-and-including ${\cal O}(M_\pi)$. The circle--cross denotes an
insertion from the next--to--leading order chiral effective
Lagrangian ${\cal L}_{\pi N}^{(2)}$.

\smallskip}
%\vskip -0.5truecm

\bigskip

Before discussing the influence of the new corrections in
eq.(\ref{a_Inew}) on the extraction of the S-wave $\pi \pi$ scattering
lengths, let us comment on the extraction of the threshold amplitudes
in ref.\cite{burkhard}. As pointed out in refs.\cite{bernard,bkmr},
only in the channels $ \pi^+ p \to \pi^+ \pi^+ n$ and $\pi^- p \to
\pi^0 \pi^0 n$ are the data  close enough to threshold to allow for an
extraction of the threshold amplitudes ${\cal A}_{10}$ and ${\cal A}_{32}$.
A global fit to all five channels as in \cite{burkhard} gives insufficient
weight to the threshold region. Correspondingly, one finds \cite{bkmr}
\beq
{\cal A}_{10} = (8.01 \pm 0.64) \, M_\pi^{-3} \, , \quad
{\cal A}_{32} = (2.53 \pm 0.14) \, M_\pi^{-3} \, ,
\label{Anew}
\eeq
which differ somewhat from the values given in \cite{burkhard}.
Ignoring for the moment $ {\cal O}(M_\pi^2)$ contributions
(i.e. setting $\tilde{d}_0 = \tilde{d}_2 =0$)
and inserting on the left hand side of eq.(\ref{a_Inew})
 the  result of the fit in eq.(\ref{Anew})
(for more details, see ref.\cite{bkmr}),
one extracts $a_0^0 (\pi \pi) = 0.23
\pm 0.02$ and $a^2_0 (\pi \pi) = -0.042 \pm 0.002 $,
 which are quite close to the CHPT prediction
at next-to-leading order. We stress, however, that these numbers
should only be considered indicative since the corrections to
eq.(\ref{a_Inew}) are not yet fully under control.

%%\newpage

Finally, we would like to make a few comments on the work of Sossi et
al. \cite{sossi}. There, the next--to--leading order $\pi \pi$
amplitude was combined with the Oset and Vicente--Vacas model \cite{oset}
and a comparison with the existing data (for $T_\pi \le 400$ MeV) was made.
This procedure is, as should be
clear from the previous discussions, not consistent since the
$\pi \pi$ and $\pi \pi N$ amplitudes should be treated at the same
order in the chiral expansion. Besides, in ref.\cite{sossi}
the mesonic  low--energy constants  based on the work
of ref.\cite{dono} are used. These are, however, determined
from a fit to $\pi \pi$ data  over an energy range which clearly
exceeds the range of validity of the one--loop calculation (see also
the discussion in section 4.1 of ref.\cite{ugm}).

\bigskip

The threshold pion production amplitude plays an important role in low
energy hadron physics because of its relationship to $\pi\pi$ scattering.
The precise nature of this relationship must be explored by an
examination of the effect of higher order ChPT contributions.
In principle, these corrections are manageable. However, the
appearance of novel counter terms  with {\it a priori} unknown
coefficients introduces uncertainties. It remains to be seen whether
the relation between the $\pi\pi$ S--wave scattering lengths and the
threshold $\pi \pi N$ amplitudes can be formulated with a sufficient
numerical accuracy to pin down $a_0^0 (\pi \pi)$ and $a_0^2 (\pi \pi)$
with an uncertainty comparable to the one of theoretical predictions
\cite{gl84}.

\section*{Acknowledgements}

The authors appreciate the hospitality of the organizers of the
{\it Chiral Dynamics: Theory and Experiment Workshop}, MIT (July 25--29,1994)
We are particularly indebted to J\"urg Gasser for many constructive remarks
and discussions. One of us (MGO) would like to acknowledge support by the U.S
Department of
Energy under Grant No.~DE-FG02-95ER40896.

%%\newpage

%%\newpage
%%%%
\end{document}